\documentstyle[twoside,fleqn,espcrc2,epsfig]{article}
\newcommand{\AmS}{{\protect\the\textfont2
  A\kern-.1667em\lower.5ex\hbox{M}\kern-.125emS}}
\def\frac#1#2{ {{#1} \over {#2} }}

\def\beq{\begin{equation}}
\def\eeq{\end{equation}}
\def\bit{\begin{itemize}}
\def\eit{\end{itemize}}
\newcommand{\ba}         {\begin{eqnarray}}
\newcommand{\ea}         {\end  {eqnarray}}
\newcommand{\ban}        {\begin{eqnarray*}}
\newcommand{\ean}        {\end  {eqnarray*}}

\def\np#1#2#3{Nucl.\ Phys.\ B#1 (19#3) #2}

\title{Numerical Stochastic Perturbation Theory.
Convergence and features of the stochastic process.
Computations at fixed (Landau) Gauge.
\vskip-2.4cm\hfill\small UPRF 99-14\vskip2.4cm}

\author{F.\ Di Renzo\address{Dipartimento di Fisica, Universit\`a di Parma
	and INFN, Gruppo Collegato di Parma, Italy}
	and L.\ Scorzato$^{\rm a}$}

\begin{document}

\begin{abstract}
Concerning Numerical Stochastic Perturbation Theory, we discuss
the convergence of the stochastic process (idea of the proof, features of
the limit distribution, rate of convergence to equilibrium).
Then we also discuss the expected fluctuations in the observables and
give some idea to reduce them. 
In the end we show that also computation of quantities at fixed
(Landau) Gauge is now possible.
\end{abstract}

\maketitle

\section{Introduction}
Numerical Stochastic Perturbation Theory (NSPT) was introduced in \cite{NSPT}
and had successful applications \cite{8l} \cite{PosterFranz}. Here we want
to describe the main features of the underlying stochastic process. In
the last part we show that a deeper understanding of the process
allows also computations at fixed (Landau) gauge.

NSPT was developed in the context of {\em Stochastic Quantization}
\cite{PW}. In this approach perturbation theory is performed through a
formal substitution of the expansion 
$U_\eta(x,t) \rightarrow \sum_k g^k U^{(k)}_\eta(x,t),$ in the
Langevin equation:
\ban
\frac{\partial}{\partial t} U_{\eta} \, = \, 
\left[ -i \nabla S[U_{\eta}] -i \eta \right] U_{\eta}
\ean
\noindent ($\nabla$ is a suitably defined group derivative). 
The Langevin equation in the algebra variables reads:
\ban
\frac{\partial}{\partial t}A^a_{\mu}(\eta,x,t) \, = \, D^{ab}_\nu
F^b_{\nu\mu}(\eta,x,t)  + \eta^a_\mu(x,t)
\ean
which gives a system of equations that can be solved numerically.

\section{Convergence and features of the stochastic process.}
 
We now consider $P_n[A^{(0)},\dots,A^{(n)},t]:$ the distribution function
of the first $n$ perturbative components of the fields at Langevin
time $t$. $P_n$ certainly exists for each $n$ and $t$ fixed. But we
need to know: whether the limit distributions exist for $t\rightarrow
\infty$, to what kind of distribution they converge and at which rate.

At this stage the answer to the first question is certainly not. This
is due to two reasons which become clear if we look at the formal
solution of the (perturbatively developed) Langevin equation in the
Fourier transformed space (with zero initial conditions):
\ba \label{Sol}
A^{(n)}_\mu(k,t) \, 
&=& \, T_{\mu\nu} \int_0^t ds \, e^{- k^2 (t-s)} f^{(n)}_\nu(k,s)
\nonumber \\
&&+ L_{\mu\nu} \int_0^t ds \, f^{(n)}_\nu(k,s)
\ea
Colour indexes are left out, $T_{\mu\nu}$ and $L_{\mu\nu}$ are the
transverse and longitudinal abelian projectors, and
$f^{(n)}_\nu(k,t) = g I^{(3) (n-1)}_\mu(k,t) + g^2 I^{(4) (n-2)}_\mu(k,t)$;
$f^{(0)}_\nu(k,t) = \eta_\nu(k,t)$. With $I^{(.)}$ we mean the 3 (4)
gluons interaction terms, which only depend on the fields till the
$n-1$ ($n-2$) perturbative order.

This process must behave like a random walk in correspondence of
both the gauge degrees of freedom and the mode $k=0$, since such degrees
of freedom have no attractive force and consequently no damping
factor. Thus they produce diverging fluctuations, even if their mean
value shall be zero since the equivalence of Stochastic and Canonical
quantization is true also when expanded in perturbation theory.

The idea is to introduce an attractive force which keep
the norms of the fields under control, without affecting the
observables \cite{Zwa}. To this end we interlace each step of Langevin
dynamic with a step of gauge transformation defined by: 
$U^{w}_{\mu}(x) = e^{w(x)}U_{\mu}(x)e^{-w(x+\mu)}$
(where $w(x) = -\lambda \sum_{\mu} \partial_{\mu} A_{\mu}$)
One can prove that, by doing this, the system gains a force that drives it
towards the Landau gauge (provided it is within the first Gribov
horizon) \cite{gf}.

We will come back to this point later when considering quantities in a
fixed gauge. For the moment suffice it to say that all gauge degree of
freedom now have an attractive force.  

\vskip .2cm

The problem of zero modes appears also in usual lattice perturbation
theory: the propagator is singular at $k=0$, and the common prescription
is to exclude the zero-momentum degree of freedom of the fields. 

In our context the problem is a bit more subtle since we work in
configuration space. Moreover if we set the $k=0$ mode to zero in the
gaussian noise $\eta^a_{\mu}(x,t)$, the free fields $A^{(0)}(x,t)$
will have no $k=0$ mode too, but this won't be true for the higher
perturbative components. In fact the interaction introduces a zero mode
contribution in the fields $A^{(p)}(x,t)$ even if it was not present in
the lower perturbative order. For instance the three gluons term gives:
\ban
\dot{A}^a_\mu(0,t)_{|3 glu} &=& \, \frac{i g f^{abc}}{2 (2 \pi)^n} 
\int dp dq \, \delta(p+q) \\
&&  A_\nu^b(-p,t) \, A_\sigma^c(-q,t) 
\, v_{\mu\nu\sigma}^{(3)}(0,p,q) 
\ean
($v_{\mu\nu\sigma}^{(3)}(k,p,q) \, = \, \delta_{\mu\nu} (k-p)_\sigma + 
\mbox{perm.}$). 
We must subtract  these 0 modes by hand at each step. 

In principle one should
subtract the zero mode component after the updating of each single
perturbative component of each single link, and before evaluating the
next perturbative order. But this would be extremely
expensive. It is convenient, instead, to subtract the zero mode after
each sweep of the whole lattice. In this way we just introduce another
error of order $\tau$, which is then extrapolated to zero.

Once implemented the two corrections described above, it is
possible to prove the convergence of the process. By that we mean that
any correlation function of any perturbative component of the fields
($\langle\prod_j A^{(n_j)}_{\mu_j}(x_j,t) \rangle$) 
has a finite limit for large Langevin time.

We just give the main ideas and results. 
We first remark that all correlations of free  
fields converge at least like $O(e^{-q^2t})$ (if $q$ is the lower
momentum). Then we proceed by induction:
It is convenient to re-write the solution (\ref{Sol}) in
discretized Langevin time ($t=N\tau$) distinguishing the memory of the
past from the new contribution of the random process:
\ban 
A^{(0)}_\mu(k,t) &=& e^{-k^2 \tau} A^{(0)}_\mu(k,t-\tau) + 
\sqrt{\tau}\eta_\mu(k,t)\\
A^{(j)}_\mu(k,t) &=& e^{-k^2 \tau} A^{(j)}_\mu(k,t-\tau) + 
\tau f^{(j)}_\mu(k,t)
\ean 
The insertion of this formula into a correlation function reduces it 
into others of lower perturbative order. 
It is not difficult to evaluate the sum of the relevant part which
survives in the limit $\tau\to 0$ at
$t=N\tau$ fixed, and then take the limit $t\to \infty$.  

The result is the following: if {\em every} degree of freedom has an
attractive force, as described above, then a limit distribution exists 
for each $P_n$, and  all their moments are finite (i.e. any
correlation function of any perturbative component of fields is
finite). Moreover  convergence  to equilibrium is damped by a factor 
$t^p e^{-k^2 t}$ where $k$ is the lower momentum contributing and $p$
is the global perturbative order of the correlation function.

\section{About fluctuations: status and ideas for improvement}
The fact that the limit distributions of these objects  produce
correlation functions which are all finite is not sufficient. We need
to have an idea of how much such correlations can grow for a high
number of fields or perturbative order. In fact some applications of
NSPT need a knowledge of the perturbative coefficient with an
extremely high precision. Since this is a stochastic method the result
is known with an error which is essentially given by the intrinsic
fluctuations of the correlation function one needs to calculate. 

To gain some insight in this problem it is useful to think of the
process in terms of the underlying gaussian process $\eta(x,t)$. Any
perturbative component $O^{(p)}$ of an observable $O$ may be seen as a
sum of correlation functions of $\eta$'s. In fact 
$\langle O^{(p)} \rangle \sim \sum_{\sigma}\langle \eta_{\sigma_1}
\ldots \eta_{\sigma_M} \rangle$
 and 
$\langle (O^{(p)})^2 \rangle \sim
\sum_{\sigma} \sum_{\pi} \langle \eta_{\pi_1} \ldots
\eta_{\pi_M} \eta_{\sigma_1} \ldots \eta_{\sigma_M} \rangle$.
The $\sigma$'s are some choices from all the possible $\eta$'s in the
process and $M$ is a number. Both depend on the theory, on the observable
and on the perturbative order $p$.  Since the fluctuations of
$O^{(p)}$ clearly depend on the number $M$ and choices $\sigma$, it
would be important to be able to say something about them.

It is quite easy to determine the number $M$ for a particular Theory
and observable. 
For instance in the $\lambda\phi^4$ theory for $O[\phi]=\phi^2$ the
relation  between the perturbative order $p$ and the maximum number of
correlated $\eta$'s $M$ is $M=2p+2$, while for the plaquette in gauge
theory the relation is simply $M=p$.

This information is widely insufficient to determine the size of the
fluctuations. There are other elements that strongly influence the
size of the fluctuations, but we can be only qualitative about them.
Consider for instance $\lambda\phi^4$: the field interacts with
itself in the same point. $\phi^{(1)} \sim (\phi^{(0)})^3$,
$\phi^{(2)} \sim (\phi^{(0)})^2\phi^{(1)} \sim (\phi^{(0)})^5$.
We have a strong contribution of correlations of the kind 
$\langle \eta^M \rangle$. We expect strong fluctuations quite soon.
The situation for $\sigma-$model is similar.
Consider instead gauge theories: Interaction is given by product of
fields of different colours and directions. $A_\mu^{a(1)} \sim g
f^{abc}A_\mu^{b(0)}A_\nu^{c(0)}$ etc. This makes the previous
phenomenon much less severe.

\noindent \underline{\em Remark:} If we want to study a fixed
observable at a fixed perturbative order we do not really need the
process $\eta$ to be gaussian. We just need a fixed number of its
correlations  to be gaussian. 
Higer moments could be chosen to be lower than those of a gaussian
distribution.
This is achieved, for instance, if one exploits combinations of
Dirac delta's: $p(x) = \sum_j w_j \delta(x-x_j)$ (see \cite{vonMises}
for the general solution).

\section{Computation of quantities at a fixed (Landau) Gauge.}
The convergence of each correlation function imply in particular that
not only gauge invariant quantities are computable but also those which
depend on the gauge.

Actually the gauge fixing procedure which is realized here is not that
of Faddeev-Popov \cite{FP} (which is possible but more expensive to
perform \cite{fermioni}), but that  introduced by Zwanziger (plus
corrections of the order of the lattice spacing and of $\tau$). 
In fact the interlaced gauge transformations described above are
equivalent to add to the Langevin equation a force
\ban
\lambda D_\mu^{ab} \partial_\nu A_\nu^b &=& - \lambda \frac{\delta
S_{GF}[A]}{\delta A_\mu^a}  
+ \lambda g f^{abc} A_\mu^b \partial_\nu A_\nu^c
\ean
(where $S_{GF}[A] = \frac{1}{2} \int dx (\partial_\nu A_\nu)^2$).
Corrections of the order of the lattice spacing are present, since the
formula above is valid only in the continuum. 
A correction of order $\tau$ is expected to come from the procedure of
interlacing a Langevin step with a gauge transformation.

Although this kind of gauge fixing is not that of Faddeev-Popov, The
Landau choice of gauge can be reproduced. There are at least two ways
of doing this. 
The first one is natural but maybe not efficient: one can perform the
calculation at different value of the ratio
$\frac{1}{\alpha}=\lambda/\tau$ and then extrapolate for large
$\frac{1}{\alpha}$. 
The second method consists in performing many gauge transformations on a
thermalized configuration. This should drive the system towards a
stationary point, where - in fact
- the Landau gauge condition is satisfied.

\end{document}